\documentclass[twocolumn,floatfix,aps,superscriptaddress]{revtex4}
\usepackage{graphicx}
\usepackage{amsmath}
\usepackage{amssymb}
\usepackage{bm}
\usepackage{stmaryrd}
\usepackage{hyperref}
 
\begin{document}

\title{Neural network decoder for topological color codes with circuit level noise}
\author{P. Baireuther}
\affiliation{Instituut-Lorentz, Universiteit Leiden, P.O. Box 9506, 2300 RA Leiden, The Netherlands}
\author{M. D. Caio}
\affiliation{Instituut-Lorentz, Universiteit Leiden, P.O. Box 9506, 2300 RA Leiden, The Netherlands}
\author{B. Criger}
\affiliation{QuTech, Delft University of Technology, P.O. Box 5046, 2600 GA Delft, The Netherlands}
\affiliation{Institute for Globally Distributed Open Research and Education (IGDORE)}
\author{C. W. J. Beenakker}
\affiliation{Instituut-Lorentz, Universiteit Leiden, P.O. Box 9506, 2300 RA Leiden, The Netherlands}
\author{T. E. O'Brien}
\affiliation{Instituut-Lorentz, Universiteit Leiden, P.O. Box 9506, 2300 RA Leiden, The Netherlands}
\date{October 2018}
\begin{abstract}
A quantum computer needs the assistance of a classical algorithm to detect and identify errors that affect encoded quantum information. At this interface of classical and quantum computing the technique of machine learning has appeared as a way to tailor such an algorithm to the specific error processes of an experiment --- without the need for \textit{a priori} knowledge of the error model. Here, we apply this technique to topological color codes. We demonstrate that a recurrent neural network with long short-term memory cells can be trained to reduce the error rate $\epsilon_{\rm L}$ of the encoded logical qubit to values much below the error rate $\epsilon_{\rm phys}$ of the physical qubits --- fitting the expected power law scaling $\epsilon_{\rm L}\propto \epsilon_{\rm phys}^{(d+1)/2}$, with $d$ the code distance. The neural network incorporates the information from ``flag qubits'' to avoid reduction in the effective code distance caused by the circuit. As a test, we apply the neural network decoder to a density-matrix based simulation of a superconducting quantum computer, demonstrating that the logical qubit has a longer life-time than the constituting physical qubits with near-term experimental parameters.
\end{abstract}
\maketitle

\section{Introduction}
\label{intro}
In fault-tolerant quantum information processing, a topological code stores the logical qubits nonlocally on a lattice of physical qubits, thereby protecting the data from local sources of noise \cite{Lid13,Ter15}. To ensure that this protection is not spoiled by logical gate operations, they should act locally. A gate where the $j$-th qubit in a code block interacts only with the $j$-th qubit of another block is called ``transversal'' \cite{Got09}. Transversal gates are desirable both because they do not propagate errors within a code block, and because they can be implemented efficiently by parallel operations.

Two families of two-dimensional (2D) topological codes have been extensively investigated, surface codes \cite{Kit03,Bra98,Fow12b,Bom07a} and color codes \cite{Bom06,Bom07}.
The two families are related: a color code is equivalent to multiple surface codes, entangled using a local unitary operation \cite{Bom12,Kub15a} that amounts to a code concatenation \cite{Cri16}.
There are significant differences between these two code families in terms of their practical implementation. 
On the one hand, the surface code has a favorably high threshold error rate for fault tolerance, but only {\sc cnot}, $X$, and $Z$ gates can be performed transversally \cite{Cam16}.
On the other hand, while the color code has a smaller threshold error rate than the surface code \cite{And11,Lan11}, it allows for the transversal implementation of the full Clifford group of quantum gates (with Hadamard, $\pi/4$ phase gate, and {\sc cnot} gate as generators) \cite{Bom15,Kub15}. While this is not yet computationally universal, it can be rendered universal using gate teleportation \cite{Got99} and magic state distillation \cite{Bra05}. Moreover, color codes are particularly suitable for topological quantum computation with Majorana qubits, since high-fidelity Clifford gates are accessible by braiding \cite{Lit17a,Lit17b}.

A drawback of color codes is that quantum error correction is more complicated than for surface codes. The identification of errors in a surface code (the ``decoding'' problem) can be mapped onto a matching problem in a graph \cite{Den02}, for which there exists an efficient solution called the ``blossom'' algorithm \cite{Edm65}. This graph-theoretic approach does not carry over to color codes, motivating the search for decoders with performance comparable to the blossom decoder, some of which use alternate graph-theoretic constructions \cite{Wan10,Duc10,Sar12,Del14,Ste14}. 

An additional complication of color codes is that the parity checks are prone to ``hook'' errors, where single-qubit errors on the ancilla qubits propagate to higher weight errors on data qubits, reducing the effective distance of the code. There exist methods due to Shor~\cite{Sho96}, Steane~\cite{Ste97}, and Knill~\cite{Kni05} to mitigate this, but these error correction methods come with much overhead because of the need for additional circuitry. An alternative scheme with reduced overhead uses dedicated ancillas (``flag qubits'') to signal the hook errors~\cite{Cha171,Cha172,Cha17,Gut18,Tan18}.

Here we show that a neural network can be trained to fault-tolerantly decode a color code with high efficiency, using only measurable data as input. No \textit{a priori} knowledge of the error model is required. Machine learning approaches have been previously shown to be successful for the family of surface codes \cite{Tor16,Var17,Bai17,Kra17,Bre17}, and applications to color codes are now being investigated \cite{Dav18,Cha18,Mas18}. We adapt the recurrent neural network of Ref.\ \onlinecite{Bai17} to decode color codes with distances up to 7, fully incorporating the information from flag qubits. A test on a density matrix-based simulator of a superconducting quantum computer \cite{Obr17} shows that the performance of the decoder is close to optimal, and would surpass the quantum memory threshold under realistic experimental conditions.

\section{Description of the problem}
\label{sec_problem}

\subsection{Color code}
\label{sec_color_code}

The color code belongs to the class of stabilizer codes \cite{Got97}, which operate by the following general scheme. We denote by $I,X,Y,Z$ the Pauli matrices on a single qubit and by $\Pi^n=\{I,X,Y,Z\}^{\otimes n}$ the Pauli group on $n$ qubits.
A set of $k$ logical qubits is encoded as a $2^k$-dimensional Hilbert space $\mathcal{H}_{L}$ across $n$ noisy physical qubits (with $2^n$-dimensional Hilbert space $\mathcal{H}_{P}$).
The logical Hilbert space is stabilized by the repeated measurement of $n-k$ parity checks $S_i\in\Pi^n$ that generate the stabilizer $\mathcal{S}(\mathcal{H}_L)$, defined as
\begin{equation}
\mathcal{S}(\mathcal{H}_L)=\{S\in\mathcal{B}(\mathcal{H}_{P}), S|\psi_L\rangle=|\psi_L\rangle\forall |\psi_L\rangle\in\mathcal{H}_L\},
\end{equation}
where $\mathcal{B}(\mathcal{H}_{P})$ is the algebra of bounded operators on the physical Hilbert space.

As errors accumulate in the physical hardware, an initial state $|\psi_L(t=0)\rangle$ may rotate out of $\mathcal{H}_L$.
Measurement of the stabilizers discretizes this rotation, either projecting $|\psi_L(t)\rangle$ back into $\mathcal{H}_L$, or into an error-detected subspace $\mathcal{H}_{\vec{s}(t)}$.
The syndrome $\vec{s}(t)\in \mathbb{Z}_2^{n-k}$ is determined by the measurement of the parity checks:
$S_i\mathcal{H}_{\vec{s}(t)}=(-1)^{s_i(t)}\mathcal{H}_{\vec{s}(t)}$.
It is the job of a classical decoder to interpret the multiple syndrome cycles and determine a correction that maps $\mathcal{H}_{\vec{s}(t)}\mapsto\mathcal{H}_L$; such decoding is successful when the combined action of error accumulation and correction leaves the system unperturbed.

This job can be split into a computationally easy task of determining a unitary that maps $\mathcal{H}_{\vec{s}(t)}\mapsto\mathcal{H}_L$ (a socalled `pure error' \cite{Pou08}), and a computationally difficult task of determining a logical operation within $\mathcal{H}_L$ to undo any unwanted logical errors.
The former task (known as `excitation removal'~\cite{Mas18}) can be performed by a `simple decoder'~\cite{Var17}.
The latter task is reduced, within the stabilizer formalism, to determining at most two parity bits per logical qubit, which is equivalent to determining the logical parity of the qubit upon measurement at time $t$~\cite{Bai17}.

\begin{figure}
\includegraphics[width=0.6\linewidth]{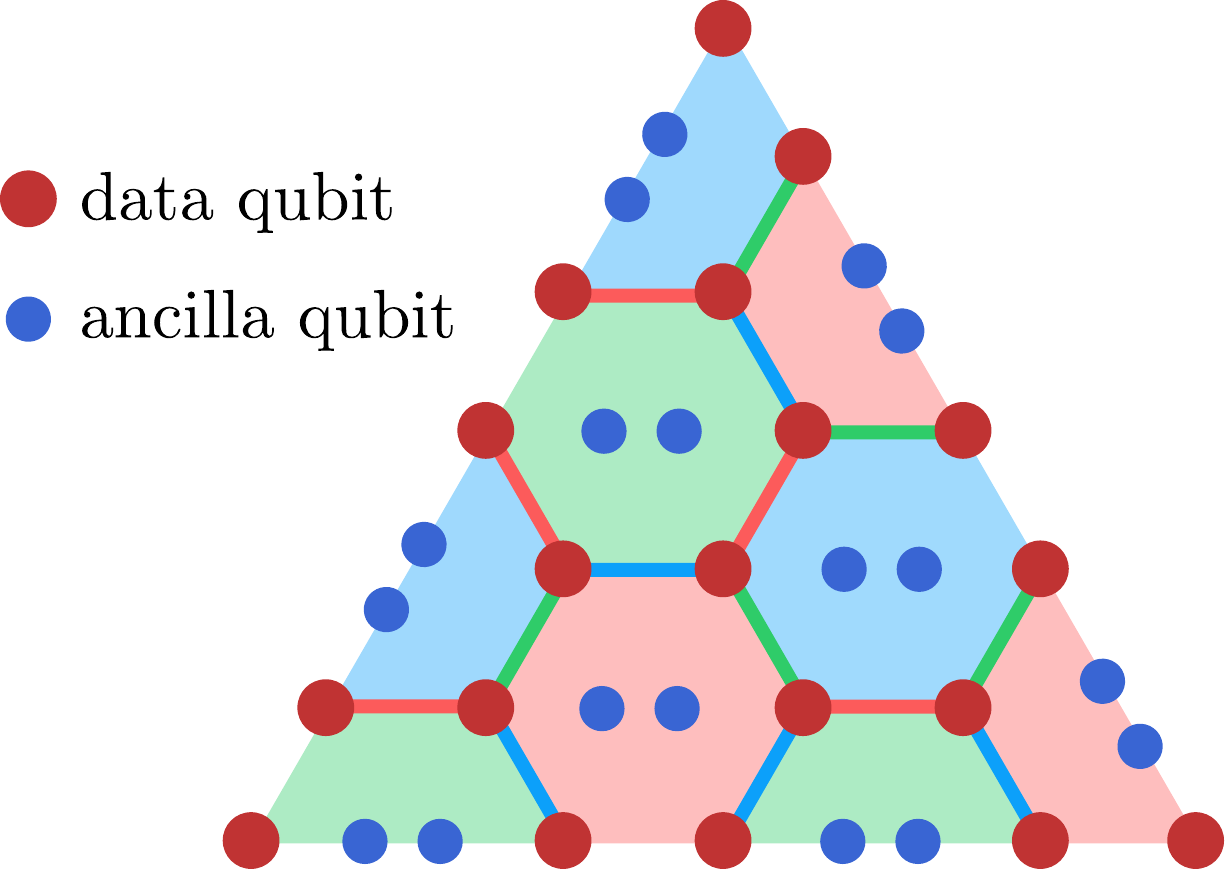}
\caption{\label{fig:color_code_schematic}Schematic layout of the distance-5 triangular color code. A hexagonal lattice inside an equilateral triangle encodes one logical qubit in 19 data qubits (one at each vertex). The code is stabilized by $6$-fold $X$ and $Z$ parity checks on the corners of each hexagon in the interior of the triangle, and $4$-fold parity checks on the boundary. For the parity checks, the data qubits are entangled with a pair of ancilla qubits inside each tile, resulting in a total of $\frac{3d^2 - 1}{2}$ qubits used to realize a distance-$d$ code. Pauli operators on the logical qubit can be performed along any side of the triangle, single-qubit Clifford operations can be applied transversally, and two-qubit joint Pauli measurements can be performed through lattice surgery to logical qubits on adjacent triangles.}
\end{figure}

We implement the color code \cite{Bom06,Bom07} on an hexagonal lattice inside a triangle, see Fig.\ \ref{fig:color_code_schematic}. (This is the 6,6,6 color code of Ref.\ \onlinecite{Lan11}.) One logical qubit is encoded by mapping vertices $v$ to data qubits $q_v$, and tiles $T$ to the stabilizers $X_T=\prod_{v\in T}X_v$, $Z_T=\prod_{v\in T}Z_v$. The simultaneous $+1$ eigenstate of all the stabilizers (the ``code space'') is twofold degenerate \cite{Kub15}, so it can be used to define a logical qubit. The number of data qubits that encodes one logical qubit is $n_{\rm data}=7$, 19, or 37 for a code with distance $d=3$, 5, or 7, respectively. (For any odd integer $d$, a distance-$d$ code can correct $(d-1)/2$ errors.) Note that $n_{\rm data}$ is less than $d^2$, being the number of data qubits used in a surface code with the same $d$ \cite{Bom07a}.

An $X$ error on a data qubit switches the parity of the surrounding $Z_T$ stabilizers, and similarly a $Z$ error switches the parity of the surrounding $X_T$ stabilizers. These parity switches are collected in the binary vector of syndrome increments $\delta\vec{s}(t)$\cite{syndrome_increment}, such that $\delta s_i=1$ signals some errors on the qubits surrounding ancilla $i$. The syndrome increments themselves are sufficient for a classical decoder to infer the errors on the physical data qubits. Parity checks are performed by entangling ancilla qubits at the center of each tile with the data qubits around the border, and then measuring the ancilla qubits (see App.~\ref{app_quantumcircuits} for the quantum circuit).

\subsection{Error model}
\label{sec_errormodel}
We consider two types of circuit-level noise models, both of which incorporate flag qubits to signal hook errors. Firstly, a simple Pauli error model allows us to develop and test the codes up to distance $d=7$. (For larger $d$ the training of the neural network becomes computationally too expensive.) Secondly, the $d=3$ code is applied to a realistic density-matrix error model derived for superconducting qubits.

In the Pauli error model, one error correction cycle of duration $t_{\rm cycle}=N_0t_{\rm step}$ consists of a sequence of $N_0=20$ steps of duration $t_{\rm step}$, in which a particular qubit is left idle, measured, or acted upon with a single-qubit rotation gate or a two-qubit conditional-phase gate. Before the first cycle we prepare all the qubits in an initial state, and we reset the ancilla qubits after each measurement.  Similarly to Ref.\ \onlinecite{Fow12b}, we allow for an error to appear at each step of the circuit and during the preparation, including the reset of the ancilla qubits, with probability $p_{\rm error}$. For the preparation errors, idle errors, or rotation errors we introduce the possibility of an $X$, $Y$, or $Z$ error with probability $p_{\rm error}/3$. Upon measurement, we record the wrong result with probability $p_{\rm error}$. Finally, after the conditional-phase gate we apply with probability $p_{\rm error}/15$ one of the following two-qubit errors: $I\otimes P$, $P\otimes I$, $P\otimes Q$, with $P,Q\in\{X,Y,Z\}$. We assume that $p_{\rm error} \ll 1$ and that all errors are independent, so that we can identify $p_{\rm error}\equiv\epsilon_{\rm phys}$ with the physical error rate per step.

The density matrix simulation uses the \emph{quantumsim} simulator of Ref.\ \onlinecite{Obr17}. We adopt the experimental parameters from that work, which match the state-of-the-art performance of superconducting transmon qubits. In the density-matrix error model the qubits are not reset between cycles of error correction. Because of this, parity checks are determined by the difference between subsequent cycles of ancilla measurement. This error model cannot be parametrized by a single error rate, and instead we compare to the decay rate of a resting, unencoded superconducting qubit. 

\subsection{Fault-tolerance}
The objective of quantum error correction is to arrive at a error rate $\epsilon_{\rm L}$ of the encoded logical qubit that is much smaller than the error rate $\epsilon_{\rm phys}$ of the constituting physical qubits. If error propagation through the syndrome measurement circuit is limited, and a ``good'' decoder is used, the logical error rate should exhibit the power law scaling~\cite{Fow12b}
\begin{equation}
\epsilon_{\rm L}=C_d \;\epsilon_{\rm phys}^{(d+1)/2},\label{epsscaling}
\end{equation}
with $C_d$ a prefactor that depends on the distance $d$ of the code but not on the physical error rate. The so-called ``pseudothreshold'' \cite{Svo05,note1}
\begin{equation}
\epsilon_{\rm{pseudo}}= \frac{1}{C_d^{2/(d-1)}} \label{pseudo_threshold}
\end{equation}
is the physical error rate below which the logical qubit can store information for a longer time than a single physical qubit.

\subsection{Flag qubits}
During the measurement of a weight-$w$ parity check with a single ancilla qubit, an error on the ancilla qubit may propagate to as many as $w/2$ errors on data qubits.
This reduces the effective distance of the code in Eq.\ \eqref{epsscaling}.
The surface code can be made resilient to such hook errors, but the color code cannot: Hook errors reduce the effective distance of the color code by a factor of two.

To avoid this degradation of the code distance, we take a similar approach to Refs.\ \onlinecite{Cha171,Cha172,Cha17,Gut18,Tan18} by adding a small number of additional ancilla qubits, socalled ``flag qubits'', to detect hook errors. For our chosen color code with weight-$6$ parity checks, we opt to use one flag qubit for each ancilla qubit used to make a stabilizer measurement. (This is a much reduced overhead in comparison to alternative approaches \cite{Sho96,Ste97,Kni05}.)
Flag and ancilla qubits are entangled during measurement and read out simultaneously (circuits given in App.~\ref{app_quantumcircuits}).
Our scheme is not \textit{a priori} fault-tolerant, as previous work has required at least $(d-1)/2$ flag qubits per stabilizer.
Instead, we rely on fitting our numeric results to Eq.~\eqref{epsscaling} with $d$ fixed to the code distance to demonstrate that our scheme is in fact fault tolerant.

\begin{figure}
\includegraphics[width=1\linewidth]{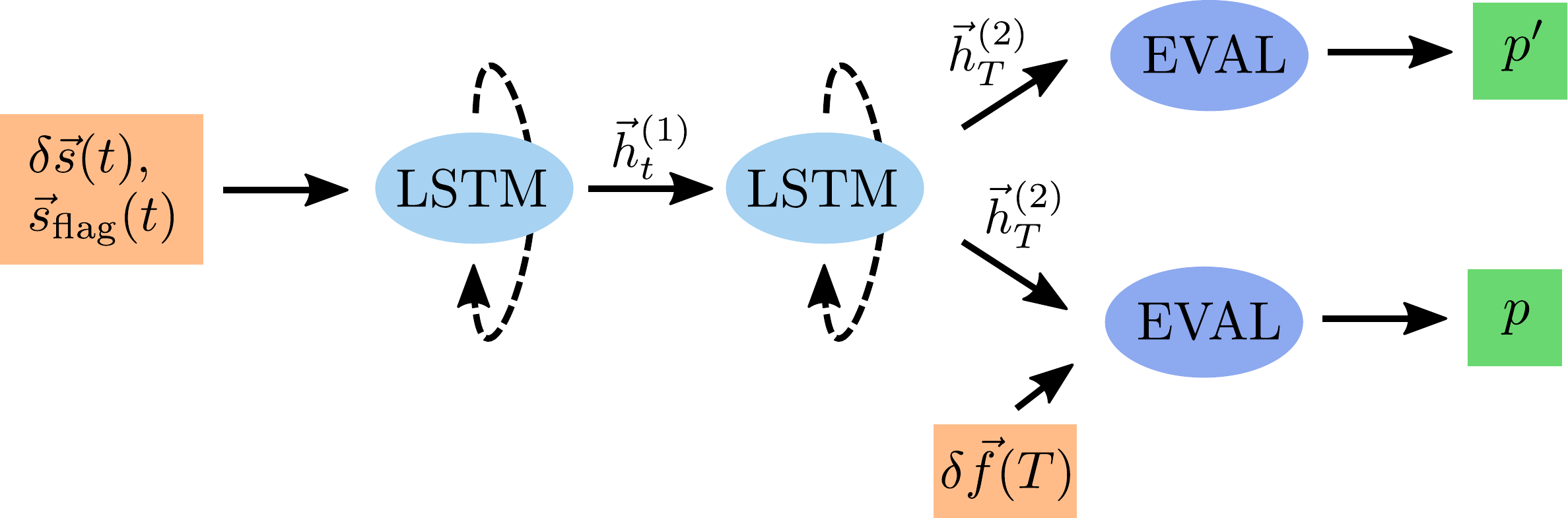}
\caption{\label{fig:network}
Architecture of the recurrent neural network decoder.
After a body of recurrent layers the network branches into two heads, each of which estimates the probability $p$ or $p'$ that the parity of bit flips at time $T$ is odd. The upper head does this solely based on syndrome increments $\delta\vec{s}$ and flag measurements $\vec{s}_{\rm flag}$ from the ancilla qubits, while the lower head additionally gets the syndrome increment $\delta\vec{f}$ from the final measurement of the data qubits. During training both heads are active, during validation and testing only the lower head is used. Ovals denote the two long short-term memory (LSTM) layers and the fully connected evaluation layers, while boxes denote input and output data. 
Solid arrows indicate data flow in the system (with $\vec{h}_t^{(1)}$ and $\vec{h}_T^{(2)}$ the output of the first and second LSTM layer), and dashed arrows indicate the internal memory flow of the LSTM layers.}
\end{figure}

\section{Neural network decoder}
\label{sec_neuralnetwork}

\subsection{Learning mechanism}
Artificial neural networks are function approximators. They span a function space that is parametrized by variables called weights and biases. The task of learning corresponds to finding a function in this function space that is close to the unknown function represented by the training data. To do this, one first defines a measure for the distance between functions and then uses an optimization algorithm to search the function space for a local minimum with respect to this measure. Finding the global minimum is in general not guaranteed, but empirically it turns out that often local minima are good approximations. For a comprehensive review see for example Refs.\ \onlinecite{Roj96, Goo16}. 

We use a specific class of neural networks known as \textit{recurrent} neural networks, where the ``function'' can represent an algorithm \cite{Sie91}. During optimization the weights and biases are adjusted such that the resulting algorithm is close to the algorithm represented by the training data.

\subsection{Decoding algorithm}
Consider a logical qubit, prepared in an arbitrary logical state $|\psi_{\rm L}\rangle$, kept for a certain time $T$, and then measured with outcome $m\in\{-1,1\}$ in the logical $Z$-basis.
Upon measurement, phase information is lost.
Hence, the only information needed in addition to $m$ is the parity of bit flips in the measurement basis. (A separate decoder is invoked for each measurement basis.)
If the bit flip parity is odd, we correct the error by negating $m \mapsto -m$. The task of decoding amounts to the estimation of the probability $p$ that the logical qubit has had an odd number of bit flips.

The experimentally accessible data for this estimation consists of measurements of ancilla and flag qubits, contained in the vectors $\delta\vec{s}(t)$ and $\vec{s}_{\rm flag}(t)$ of syndrome increments and flag measurements, and, at the end of the experiment, the readout of the data qubits. From this data qubit readout a final syndrome increment vector $\delta \vec{f}(T)$ can be calculated. Depending on the measurement basis, it will only contain the $X$ or the $Z$ stabilizers. 

Additionally, we also need to know the true bit flip parity $p_{\rm true}$. To obtain this we initialize the logical qubit at $|\psi_{\rm L}\rangle \equiv |0\rangle$ ($|\psi_L\rangle \equiv |1\rangle$ would be an equivalent choice) and then compare the final measured logical state to this initial logical state to obtain the true bit flip parity $p_{\rm true} \in \{0, 1\}.$

An efficient decoder must be able to decode an arbitrary and unspecified number of error correction cycles.
As a feedforward neural network requires a fixed input size, it is impractical to train such a neural network to decode the entire syndrome data in a single step, as this would require a new network (and new training data) for every experiment with a different number of cycles.
Instead, a neural network for quantum error correction must be cycle-based: It must be able to parse repeated input of small pieces of data (e.g.~syndrome data from a single cycle) until called upon by the user to provide output.
Importantly, this requires the decoder to be translationally invariant in time: It must decode late rounds of syndrome data just as well as the early rounds.
To achieve this, we follow Ref.\ \onlinecite{Bai17} and use a recurrent neural network of long short-term memory (LSTM) layers~\cite{Hoc97} --- with one significant modification, which we now describe.

The time-translation invariance of the error propagation holds for the ancilla qubits, but it is broken by the final measurement of the data qubits --- since any error in these qubits will not propagate forward in time. To extract the time-translation invariant part of the training data, in Ref.\ \onlinecite{Bai17} two separate networks were trained in parallel, one with and one without the final measurement input. Here, we instead use a single network with two heads, as illustrated in Fig.\ \ref{fig:network}. The upper head sees only the translationally invariant data, while the lower head solves the full decoding problem. In appendix \ \ref{details_appendix} we describe the details of the implementation.

The switch from two parallel networks to a single network with two heads offers several advantages:
(1) The number of LSTM layers and the computational cost is cut in half;
(2) The network can be trained on a single large error rate, then used for smaller error rates without retraining;
(3) The bit flip probability from the upper head provides a so-called Pauli frame decoder \cite{Ter15}.

In the training stage the bit flip probabilities $p'$ and $p$ $\in[0,1]$ from the upper and lower head are compared with the true bit flip parity $p_{\rm true}\in \{0,1\}$. By adjusting the weights of the network connections a cost function is minimized in order to bring $p',p$ close to $p_{\rm true}$. We carry out this machine learning procedure using the \textit{TensorFlow} library \cite{tensorflow}.

After the training of the neural network has been completed we test the decoder on a fresh dataset. Only the lower head is active during the testing stage. If the output probability $p<0.5$, the parity of bit flip errors is predicted to be even and otherwise odd. We then compare this to $p_{\rm true}$ and average over the test dataset to obtain the logical fidelity ${\cal F}(t)$. Using a two-parameter fit to \cite{Obr17}
\begin{equation}
\mathcal{F}(t)=\tfrac{1}{2}+\tfrac{1}{2}(1-2\epsilon_{\rm L})^{(t-t_0)/t_{\rm step}},\label{eq:fidelity}
\end{equation}
we determine the logical error rate $\epsilon_{\rm L}$ per step of the decoder.

\section{Neural network performance}
\label{sec_performance}

\subsection{Power law scaling of the logical error rate}

\begin{figure}
\includegraphics[width=\columnwidth]{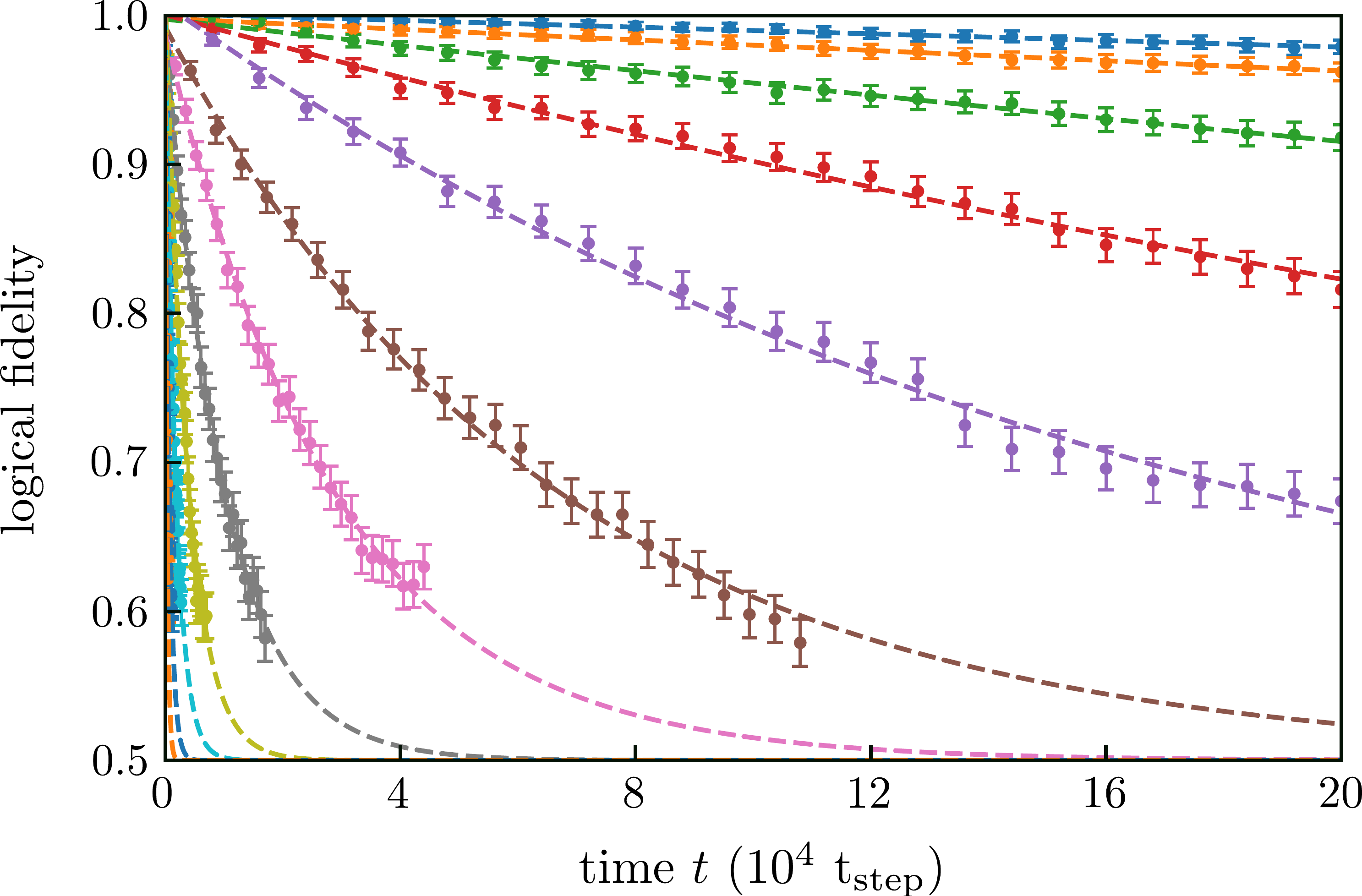}
\caption{\label{fig:decay_curves_dist3} Decay of the logical fidelity for a distance-3 color code. The curves correspond to different physical error rates $\epsilon_{\rm phys}$ per step, from top to bottom:
$1.6 \cdot 10^{-5}$, $2.5 \cdot 10^{-5}$, $4.0 \cdot 10^{-5}$, $6.3 \cdot 10^{-5}$, $1.0\cdot10^{-4}$, $1.6 \cdot 10^{-4}$, $2.5 \cdot 10^{-4}$, $4.0 \cdot 10^{-4}$, $6.3 \cdot 10^{-4}$, $1.0\cdot10^{-3}$,  $1.6 \cdot 10^{-3}$, $2.5 \cdot 10^{-3}$.
Each point is averaged over $10^3$ samples. Error bars are obtained by bootstrapping. Dashed lines are two-parameter fits to Eq.\ \eqref{eq:fidelity}.}
\end{figure}

\begin{figure} 
\includegraphics[width=\columnwidth]{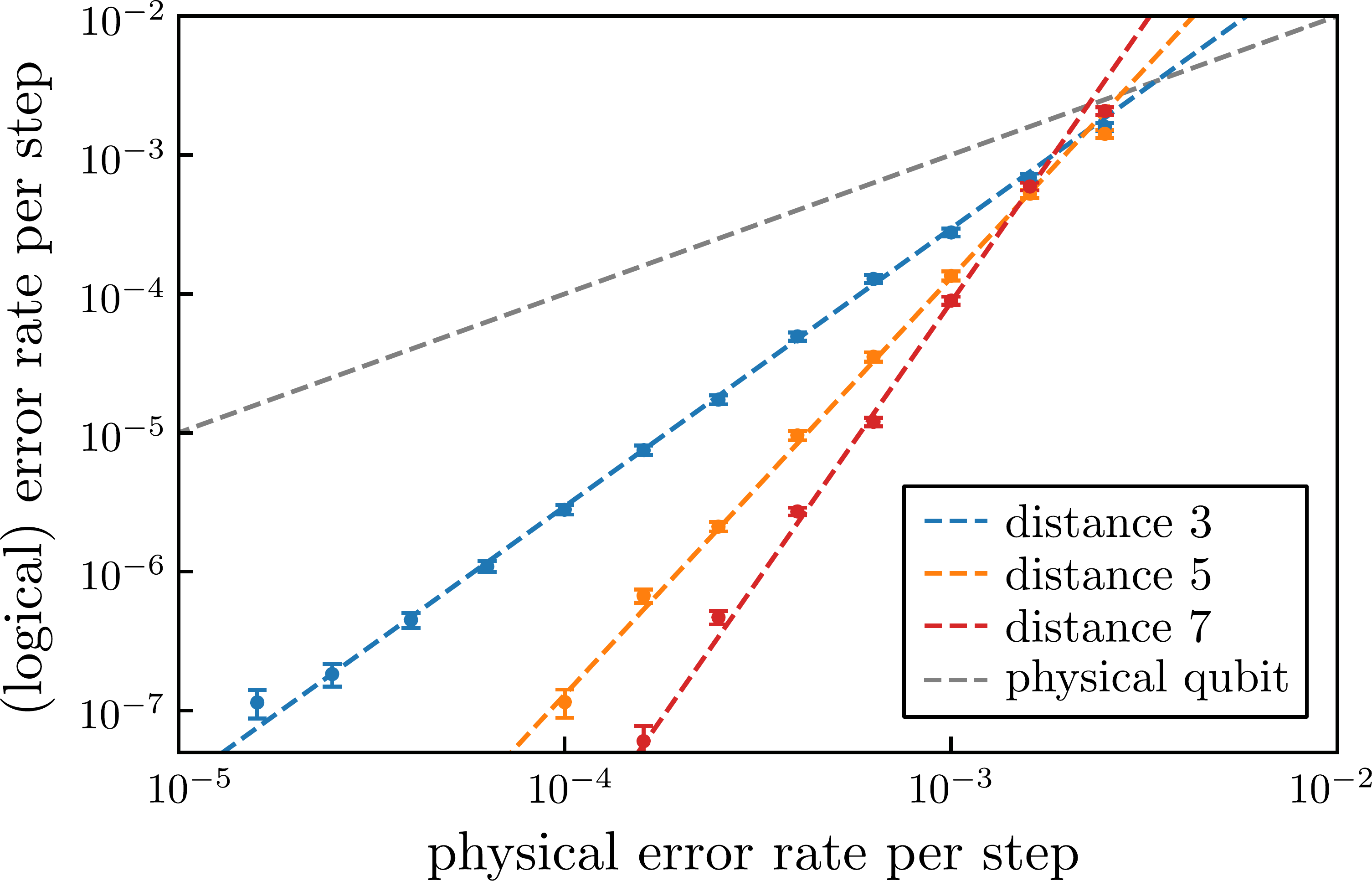}
\caption{\label{fig:benchmark} In color: Log-log plot of the logical versus physical error rates per step, for distances $d=3,5,7$ of the color code. The dashed line through the data points has the slope given by Eq.\ \eqref{epsscaling}. Quality of fit indicates that at least $\left \lfloor \tfrac{1}{2}(d+1) \right \rfloor$ independent physical errors must occur in a round to generate a logical error in that round, so syndrome extraction is fault-tolerant. In gray: Error rate of a single physical (unencoded) qubit. The error rates at which this line intersects with the lines for the encoded qubits are the pseudothresholds.
}
\end{figure} 

Results for the distance-3 color code are shown in Fig.\ \ref{fig:decay_curves_dist3} (with similar plots for distance-5 and distance-7 codes in App.~\ref{sec_57}). These results demonstrate that the neural network decoder is able to decode a large number of consecutive error correction cycles. The dashed lines are fits to Eq.\ \eqref{eq:fidelity}, which allow us to extract the logical error rate $\epsilon_{\rm L}$ per step, for different physical error rates $\epsilon_{\rm phys}$ per step.

Figure \ref{fig:benchmark} shows that the neural network decoder follows a power law scaling \eqref{epsscaling} with $d$ fixed to the code distance. This shows that the decoder, once trained using a single error rate, operates equally efficiently when the error rate is varied, and that our flag error correction scheme is indeed fault-tolerant. The corresponding pseudothresholds \eqref{pseudo_threshold} are listed in Table \ref{tb:pseudo_thresholds}.

\begin{table}[h]
\begin{tabular}{c|c}
 
  distance $d$ & pseudothreshold $\epsilon_{\rm pseudo}$ \\
  \hline
  3 & 0.0034 \\
  \hline
  5 & 0.0028 \\
  \hline
  7 & 0.0023 
\end{tabular}
\caption{Pseudothresholds calculated from the data of Fig.\ \ref{fig:benchmark}, giving the physical error rate below which the logical qubit can store information for a longer time than a single physical qubit.}
\label{tb:pseudo_thresholds}
\end{table}

\subsection{Performance on realistic data}

\begin{figure}
\includegraphics[width=\columnwidth]{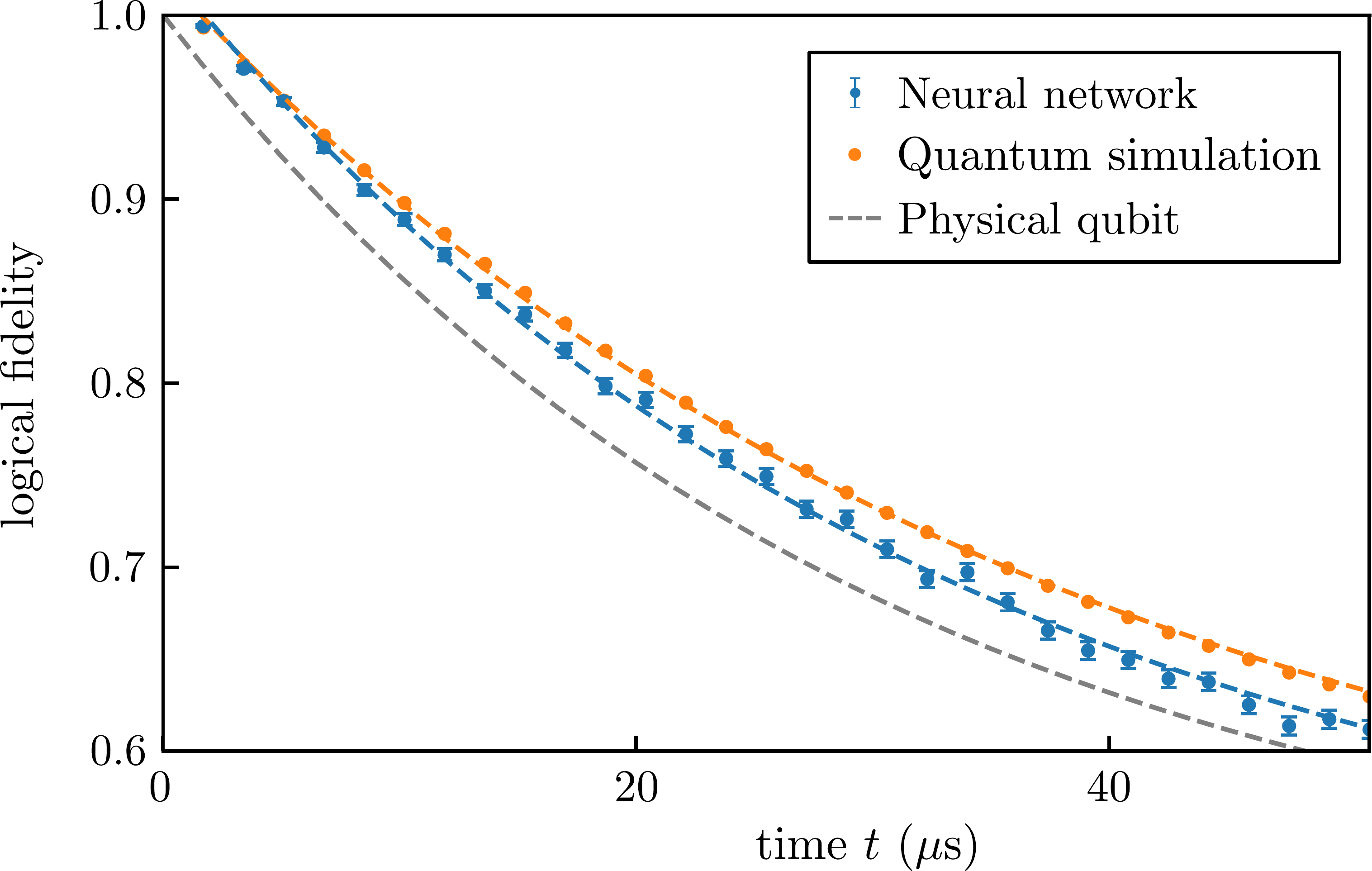}
\caption{\label{fig:quantumsim}Same as Fig.\ \ref{fig:decay_curves_dist3}, but for a density matrix-based simulation of an array of superconducting transmon qubits. Each point is an average over $10^4$ samples. The density matrix-based simulation gives the performance of an optimal decoder, with a logical error rate $\epsilon_{\rm optimal}=0.0132$ per microsecond. From this, and the error rate $\epsilon_{\rm L}=0.0148$ per microsecond obtained by the neural network, we calculate the neural network decoder efficiency to be 0.89. The average fidelity of an unencoded transmon qubit at rest with the same physical parameters is plotted in gray.
}
\end{figure}

To assess the performance of the decoder in a realistic setting, we have implemented the distance-3 color code using a density matrix based simulator of superconducting transmon qubits \cite{Obr17}. We have then trained and tested the neural network decoder on data from this simulation. In Fig.\ \ref{fig:quantumsim} we compare the decay of the fidelity of the logical qubit as it results from the neural network decoder with the fidelity extracted from the simulation \cite{Obr17}. The latter fidelity determines via Eq.\ \eqref{eq:fidelity} the logical error rate $\epsilon_{\rm optimal}$ of an optimal decoder. For the distance-3 code we find $\epsilon_{\rm L}=0.0148$ and $\epsilon_{\rm optimal}=0.0132$ per microsecond. This can be used to calculate the decoder efficiency \cite{Obr17} $\epsilon_{\rm optimal}/ \epsilon_{\rm L}=0.89$, which measures the performance of the neural network decoder separate from uncorrectable errors. The dashed gray line is the average fidelity (following Eq.~\eqref{eq:fidelity}) of a single physical qubit at rest, corresponding to an error rate of $0.0164$~\cite{Obr17}. This demonstrates that, even with realistic experimental parameters, a logical qubit encoded with the color code has a longer life-time than a physical qubit.

\section{Conclusion}
\label{sec_conclusion}
We have presented a machine-learning based approach to quantum error correction for the topological color code. We believe that this approach to fault-tolerant quantum computation can be used efficiently in experiments on near-term quantum devices with relatively high physical error rates (so that the neural network can be trained with relatively small datasets). In support of this, we have presented a density matrix simulation \cite{Obr17} of superconducting transmon qubits (Fig.~\ref{fig:quantumsim}), where we obtain a decoder efficiency of $\eta_d=0.89$. 

Independently of our investigation, three recent works have shown how a neural network can be applied to color code decoding. Refs.~\onlinecite{Dav18} and \onlinecite{Mas18} only consider single rounds of error correction, and cannot be extended to a multi-round experiment or circuit-level noise. Ref.\ \onlinecite{Cha18} uses the Steane and Knill error correction schemes when considering color codes, which are also fault-tolerant against circuit-level noise, but have larger physical qubit requirements than flag error correction. None of these works includes a test on a simulation of physical hardware.

\acknowledgments

We have benefited from discussions with Christopher Chamberland, Andrew Landahl, Daniel Litinski, and Barbara Terhal. This research was supported by the Netherlands Organization for Scientific Research (NWO/OCW) and by an ERC Synergy Grant.

\appendix

\begin{figure*}
\includegraphics[width=0.6\textwidth]{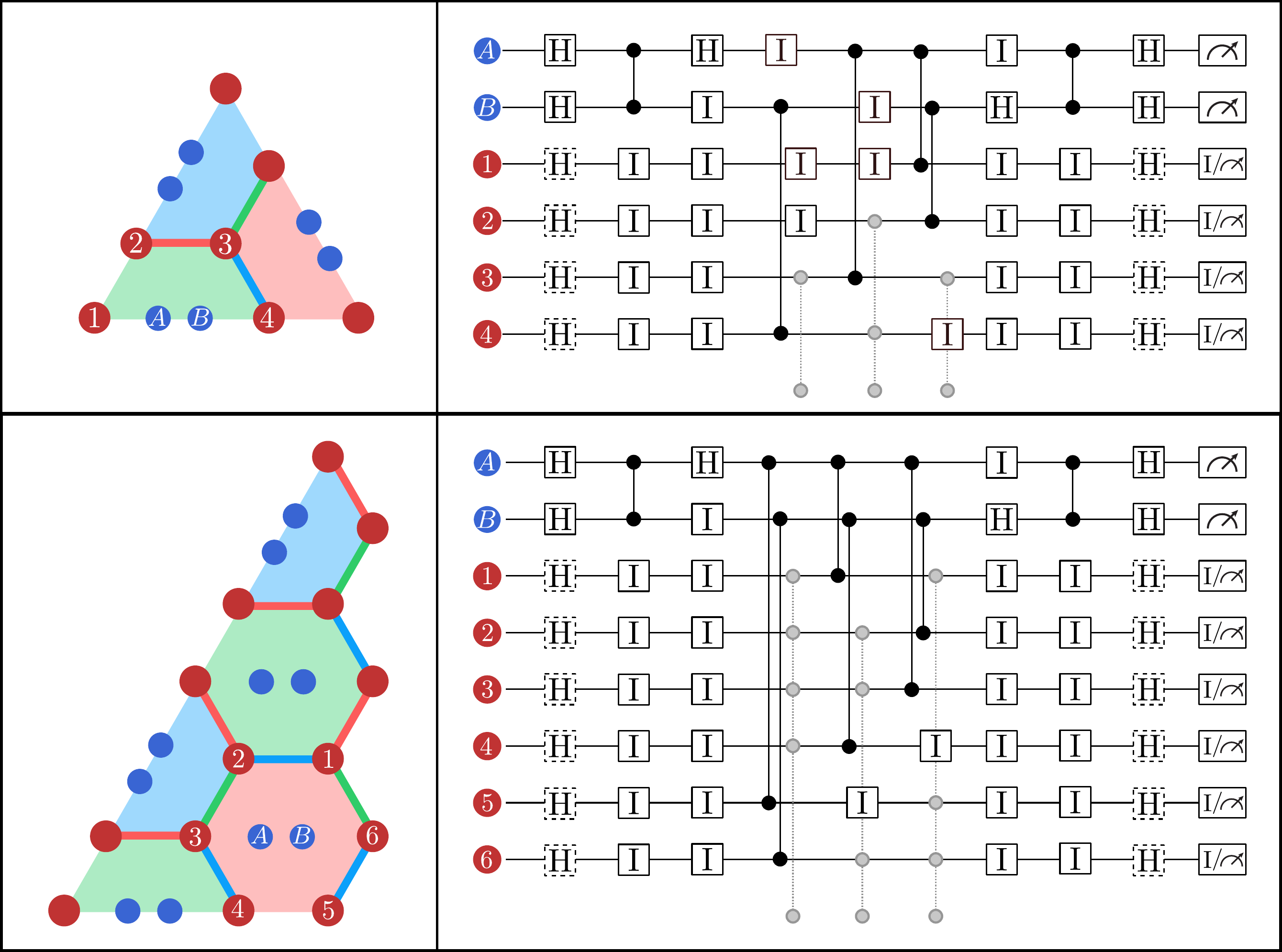}
\caption{\label{fig:circuit}
Top left: Schematic of a 6-6-6 color code with distance 3.
Top right: Stabilizer measurement circuits for a plaquette on the boundary. 
Bottom left: Partial schematic of a 6-6-6 color code with distance larger than 3. Bottom right: Stabilizer measurement circuits for a plaquette in the bulk. For the circuits in the right panels, the dashed Hadamard gates are only present when measuring the $X$ stabilizers, and are replaced by idling gates for the $Z$ stabilizer circuits; the grayed out gates correspond to conditional-phase gates between the considered data qubits and ancillas belonging to other plaquettes; and the data qubits are only measured after the last round of error correction, otherwise they idle whilst the ancillas are measured.}
\end{figure*}

\section{Quantum circuits}
\label{app_quantumcircuits}

\subsection{Circuits for the Pauli error model}

Fig.\ \ref{fig:circuit} shows the circuits for the measurements of the $X$ and $Z$ stabilizers in the Pauli error model. To each stabilizer, measured with the aid of an ancilla qubit, we associate a second ``flag'' ancilla qubit with the task of spotting faults of the first ancilla \cite{Cha171,Cha172,Cha17,Gut18,Tan18}. This avoids hook errors (errors that propagate from a single ancilla qubit onto two data qubits), which would reduce the distance of the code.
After the measurement of the $X$ stabilizers, all the ancillas are reset to $|0\rangle$ and reused for the measurement of the $Z$ stabilizers. Before finally measuring the data qubits, we allow the circuit to run for $T$ cycles.

\subsection{Measurement processing for the density-matrix error model}\label{app:measurement_processing}

For the density matrix simulation, neither ancilla qubits nor flag qubits are reset between cycles, leading to a more involved extraction process of both $\delta\vec{s}(t)$ and $\vec{s}_{\rm flag}(t)$, as we now explain.

Let $\vec{m}(t)$ and $\vec{m}_{\rm flag}(t)$ be the actual ancilla and flag qubit measurements taken in cycle $t$, and $\vec{m}^0(t)$, $\vec{m}^0_{\rm flag}(t)$ be compensation vectors of ancilla and flag measurements that would have been observed had no errors occurred in this cycle.
Then, 
\begin{align}
\delta\vec{s}(t)&=\vec{m}(t)+\vec{m}^0(t)\mod 2,\label{eq:stabinc}\\ \vec{s}_{\rm flag}(t)&=\vec{m}_{\rm flag}(t)+\vec{m}_{\rm flag}^0(t)\mod 2.
\end{align}
Calculation of the compensation vectors $\vec{m}^0(t)$ and $\vec{m}_{\rm flag}^0(t)$ requires knowledge of the stabilizer $\vec{s}(t-1)$, and the initialization of the ancilla qubits $\vec{m}(t-1)$ and the flag qubits $\vec{m}_{\rm flag}(t-1)$, being the combination of the effects of individual non-zero terms in each of these.

Note that a flag qubit being initialized in $|1\rangle$ will cause errors to propagate onto nearby data qubits, but these errors can be predicted and removed prior to decoding with the neural network.
In particular, let us concatenate $\vec{m}(t)$, $\vec{m}_{\rm flag}(t)$ and $\vec{s}(t)$ to form a vector $\vec{d}(t)$.
The update may then be written as a matrix multiplication:
\begin{equation}
\vec{m}_{\rm flag}^0(t)=M_f\vec{d}(t-1)\mod 2,
\end{equation}
Where $M_f$ is a sparse, binary matrix.
The syndromes $\vec{s}(t)$ may be updated in a similar fashion
\begin{equation}
\vec{s}(t)=\vec{s}(t-1)+\delta\vec{s}(t)+M_s\vec{d}(t-1)\; \mod \; 2,
\end{equation}
where $M_s$ is likewise sparse.
Both $M_f$ and $M_s$ may be constructed by modeling the stabilizer measurement circuit in the absence of errors.
The sparsity in both matrices reflect the connectivity between data and ancilla qubits; for a topological code, both $M_f$ and $M_s$ are local.
The calculation of the syndrome increments $\delta\vec{s}(t)$ via Eq.\ \eqref{eq:stabinc} does not require prior calculation of $\vec{s}(t)$.

\section{Details of the neural network decoder}
\label{details_appendix}

\subsection{Architecture}
The decoder consists of a double headed network, see Fig.\ \ref{fig:network}, which we implement using the \textit{TensorFlow} library \cite{tensorflow}. The source code of the neural network decoder can be found at~\cite{Decoder}. The network maps a list of syndrome increments $\delta \vec{s}(t)$ and flag measurements $\vec{s}_{\rm flag}(t)$ with $t/t_{\rm cycle}=1, 2, ..., T$ to a pair of probabilities $p',p\in [0,1]$. (In what follows we measure time in units of the cycle duration $t_{\rm cycle}=N_0 t_{\rm step}$, with $N_0=20$.) The lower head gets as additional input a single final syndrome increment $\delta \vec f(T)$. The cost function $I$ that we seek to minimize by varying the weight matrices $\bm w$ and bias vectors $\vec{b}$ of the network is the cross-entropy
\begin{equation}
H(p_1, p_2) = -p_1 \log p_2 - (1-p_1) \log(1-p_2)
\end{equation}
 between these output probabilities and the true final parity $p_{\rm true}\in\{0,1\}$ of bit flip errors:
\begin{align}
I ={}& H(p_{\rm true}, p) + \tfrac{1}{2} H(p_{\rm true}, p') + c ||{\bm w_{\rm EVAL}}||^2. \label{eq:cost_functionapp}
\end{align}
The term $c ||{\bm w_{\rm EVAL}}||^2$ with $c \ll 1$ is a regularizer, where ${\bm w_{\rm EVAL}} \subset {\bm w}$ are the weights of the evaluation layer.

The body of the double headed network is a recurrent neural network, consisting of two LSTM layers ~\cite{Hoc97, Ger00, Zar14}. Each of the LSTM layers has two internal states, representing the long-term memory $\vec{c}_t^{(i)} \in \mathbb R^N$ and the short-term memory $\vec{h}_t^{(i)} \in \mathbb R^N$, where $N=32, 64, 128$ for distances $d=3, 5, 7$. Internally, an LSTM layer consists of four simple neural networks that control how the short- and long-term memory are updated based on their current states and new input $x_t$. Mathematically, it is described by the following equations \cite{Ger00, Zar14}:
\begin{subequations}
\begin{align}
  \vec{i}_t &= \sigma({\bm w_{i}} \vec{x}_t + {\bm v_{i}} \vec{h}_{t-1} + \vec{b}_i), \\
  \vec{f}_t &= \sigma({\bm w_{f}} \vec{x}_t + {\bm v_{f}} \vec{h}_{t-1} + \vec{b}_f), \\
  \vec{o}_t &= \sigma({\bm w_{o}} \vec{x}_t + {\bm v_{o}} \vec{h}_{t-1} + \vec{b}_o), \\
  \vec{m}_t &=  \tanh({\bm w_{m}} \vec{x}_t + {\bm v_{m}} \vec{h}_{t-1} + \vec{b}_m), \\
  \vec{c}_t &= \vec{f}_t \odot \vec{c}_{t-1} + \vec{i}_t \odot \vec{m}_t \\
  \vec{h}_t &= \vec{o}_t \odot \tanh(\vec{c}_t).
\end{align}
\end{subequations}
Here ${\bm w}$ and ${\bm v}$ are weight matrices, $\vec{b}$ are bias vectors, $\sigma$ is the sigmoid function, and $\odot$ is the element-wise product between two vectors. The letters $i$, $m$, $f$, and $o$ label the four internal neural network gates: input, input modulation, forget, and output. The first LSTM layer gets the syndrome increments $\delta \vec{s}(t)$ and flag measurements $\vec{s}_{\rm flag}(t)$ as input, and outputs its short term memory states $\vec{h}_t^{(1)}$. These states are in turn the input to the second LSTM layer.

The heads of the network consist of a single layer of rectified linear units, whose outputs are mapped onto a single probability using a sigmoid activation function. The input of the two heads is the last short-term memory state of the second LSTM layer, subject to a rectified linear activation function ${\rm{ReL}}(\vec{h}_T^{(2)})$. For the lower head we concatenate ${\rm{ReL}}(\vec{h}_T^{(2)})$ with the final syndrome increment $\delta \vec{f}(T)$.

\subsection{Training and evaluation}
\label{app:train_and_eval}

\begin{figure}
\includegraphics[width=\columnwidth]{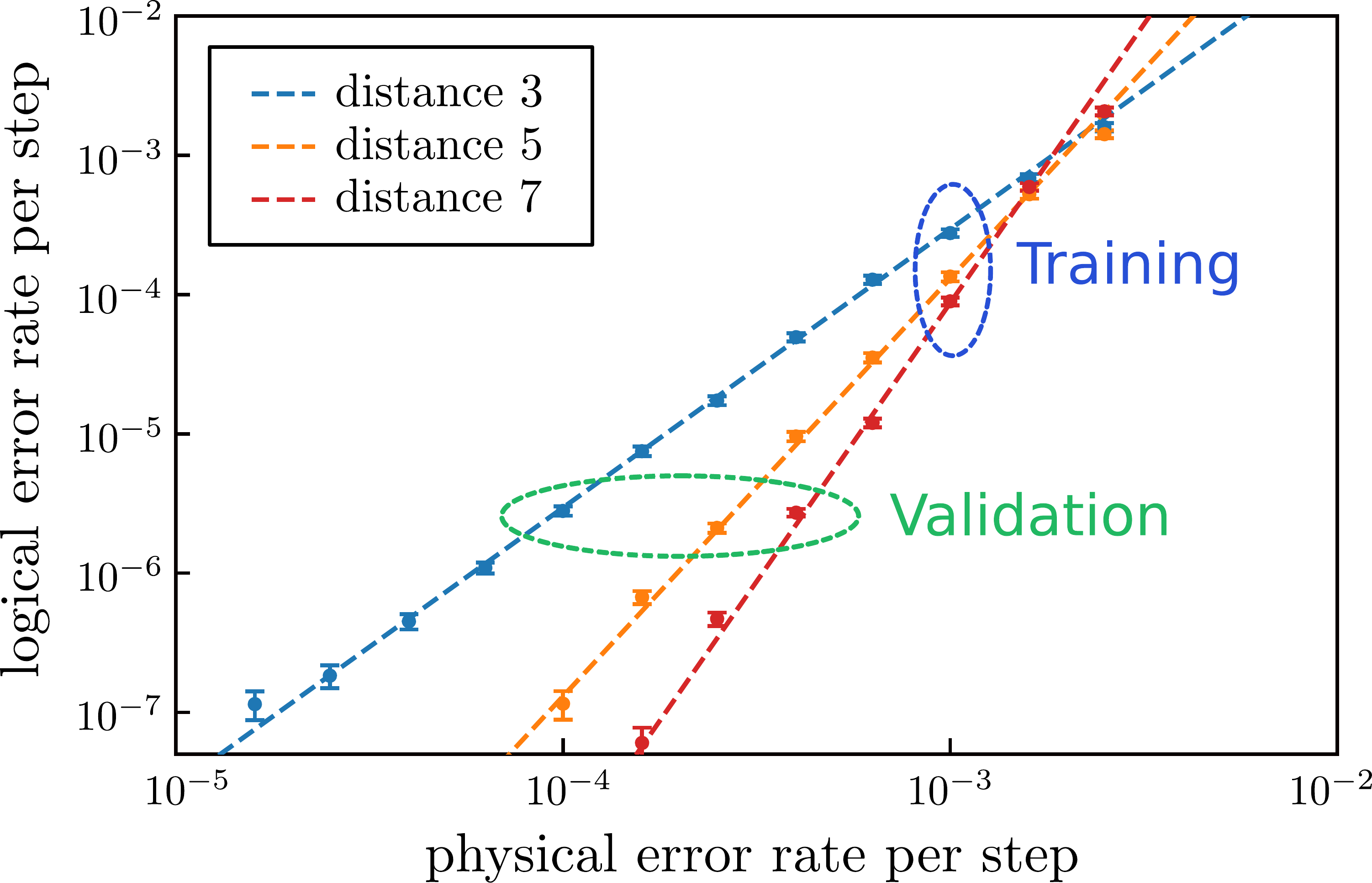}
\caption{\label{fig:benchmark_details}Same as Fig.\ \ref{fig:benchmark}. The blue ellipse indicates the error rates used during training, and the green ellipse indicates the error rates used for validation.}
\end{figure}

We use three separate datasets for each code distance. The training dataset is used by the optimizer to optimize the trainable variables of the network. It consists of $2\cdot 10^6$ sequences of lengths between $T=1$ and $T=40$ at a large error rate of $p=10^{-3}$ for distances $3$ and $5$, and of $5\cdot 10^6$ sequences for distance $7$. At the end of each sequence, it contains the final syndrome increment $\delta \vec{f}(T)$ and the final parity of bit flip errors $p_{\rm true}$. After each training epoch, consisting of $3000$ to $5000$ mini-batches of size $64$, we validate the network (using only the lower head) on a validation dataset consisting of $10^3$ sequences of $30$ different lengths between $1$ and $10^4$ cycles. By validating on sequences much longer than the sequences in the training dataset, we select the instance of the decoder that generalizes best to long sequences. The error rates of the validation datasets are chosen such that they are the largest error rate for which the expected logical fidelity after $10^4$ cycles is still larger than $0.6$ (see Fig.\ \ref{fig:benchmark_details}), because if the logical fidelity approaches $0.5$ a meaningful prediction is no longer possible. The error rates of the validation datasets are $1\cdot 10^{-4}$, $2.5 \cdot 10^{-4}$, $4 \cdot 10^{-4}$ for distances $3$, $5$, $7$ respectively. To avoid unproductive fits during the early training stages, we calculate the logical error rate with a single parameter fit to Eq.\ \eqref{eq:fidelity} by setting $t_0=0$ during validation. If the logical error rate reaches a new minimum on the validation dataset, we store this instance of the network.

We stop the training after $10^3$ epochs. One training epoch takes about one minute for distance 3 (network size $N=32$) when training on sequences up to length $T=20$ and about two minutes for sequences up to length $T=40$ on an Intel(R) Xeon(R) CPU E3-1270 v5 @ 3.60GHz. For distance 5 ($N=64$, $T=1, 2, ..., 40$) one epoch takes about five minutes and for distance 7 ($N=128$, $T=1, 2, ..., 40$) about ten minutes.

To keep the computational effort of the data generation tractable, for the density matrix-based simulation (Fig.\ \ref{fig:quantumsim}) we only train on $10^6$ sequences of lengths between $T=1$ and $T=20$ cycles and validate on $10^4$ sequences of lengths between $T=1$ and $T=30$ cycles. For the density matrix-based simulation, all datasets have the same error rate.

We train using the Adam optimizer \cite{Kin14} with a learning rate of $10^{-3}$. To avoid over-fitting and reach a better generalization of the network to unseen data, we employ two additional regularization methods: Dropout and weight regularization. Dropout with a keep probability of $0.8$ is applied to the output of each LSTM layer and to the output of the hidden units of the evaluation layers. Weight regularization, with a prefactor of $c=10^{-5}$, is only applied to the weights of the evaluation layers, but not to the biases. The hyperparameters for training rate, dropout, and weight regularization were taken from \cite{Bai17}. The network sizes were chosen by try and error to be as small as possible without fine-tuning, restricted to powers of two $N=2^n$.

After training is complete we evaluate the decoder on a test dataset consisting of $10^3$ ($10^4$ for the density matrix-based simulation) sequences of lengths such that the logical fidelity decays to approximately $0.6$, but no more than $T=10^4$ cycles. Unlike for the training and validation datasets, for the test dataset we sample a final syndrome increment and the corresponding final parity of bit flip errors after each cycle. We then select an evenly distributed subset of $t_n= n \Delta T < T_{\text{max}}$ cycles, where $\Delta T$ is the smallest integer for which the total number of points is less than 50, for evaluation. This is done in order to reduce the needed computational resources. The logical error rate $\epsilon$ per step is determined by a fit of the fidelity to Eq.\ \eqref{eq:fidelity}.

\subsection{Pauli frame updater}

We operate the neural network as a bit-flip decoder, but we could have alternatively operated it as a Pauli frame updater. We briefly discuss the connection between the two modes of operation.

Generally, a decoder executes a classical algorithm that determines the operator $P(t)\in\Pi^n$ (the so-called Pauli frame) which transforms $|\psi_L(t)\rangle$ back into the logical qubit space $\mathcal{H}_{\vec{0}}=\mathcal{H}_{L}$.
Equivalently (with minimal overhead), a decoder may keep track of logical parity bits $\vec{p}$ that determine whether the Pauli frame of a `simple decoder'~\cite{Var17} commutes with a set of chosen logical operators for each logical qubit.

The second approach of bit-flip decoding has two advantages over Pauli frame updates:
Firstly, it removes the gauge degree of freedom of the Pauli frame ($SP(t)$ is an equivalent Pauli frame for any stabilizer $S$).
Secondly, the logical parity can be measured in an experiment, where no `true' Pauli frame exists (due to the gauge degree of freedom).

Note that in the scheme where flag qubits are used without reset, the errors from qubits initialized in $\left \vert 1 \right \rangle$ may be removed by the simple decoder without any additional input required by the neural network.

\section{Results for distance-5 and distance-7 codes}
\label{sec_57}

\begin{figure}[ht!]
\includegraphics[width=\columnwidth]{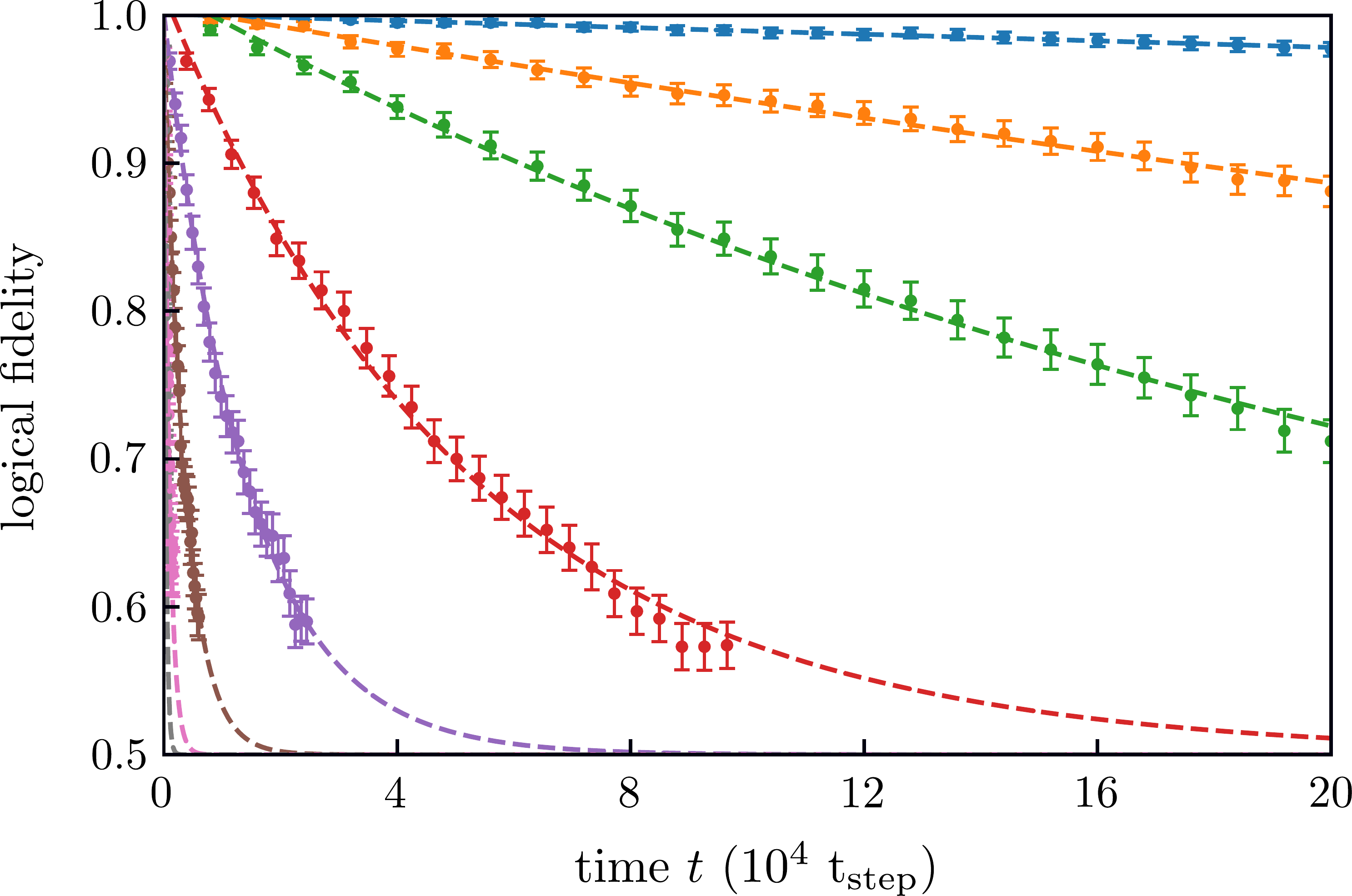}
\caption{\label{fig:decay_curves_dist5}Same as Fig.\ \ref{fig:decay_curves_dist3} for a distance-5 code; the physical error rate $\epsilon_{\rm phys}$ from top to bottom is: $1.0\cdot10^{-4}$, $1.6 \cdot 10^{-4}$, $2.5 \cdot 10^{-4}$, $4.0 \cdot 10^{-4}$, $6.3 \cdot 10^{-4}$, $1.0\cdot10^{-3}$,  $1.6 \cdot 10^{-3}$, $2.5 \cdot 10^{-3}$.}
\end{figure}

\begin{figure}
\includegraphics[width=\columnwidth]{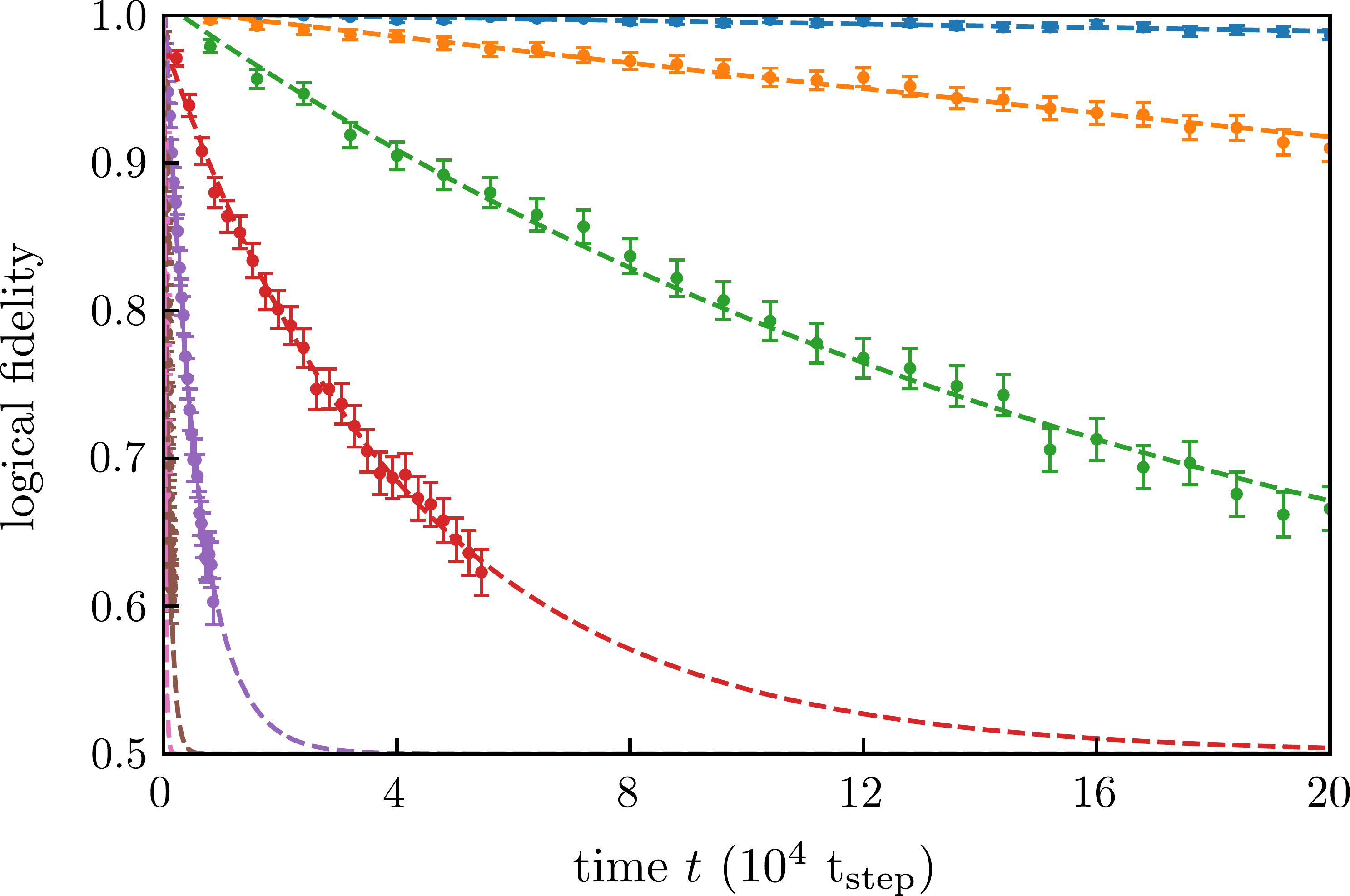}
\caption{\label{fig:decay_curves_dist7}Same as Fig.\ \ref{fig:decay_curves_dist3} for a distance-7 code; the physical error rate $\epsilon_{\rm phys}$ from top to bottom is: $1.6 \cdot 10^{-4}$, $2.5 \cdot 10^{-4}$, $4.0 \cdot 10^{-4}$, $6.3 \cdot 10^{-4}$, $1.0\cdot 10^{-3}$,  $1.6 \cdot 10^{-3}$, $2.5 \cdot 10^{-3}$.}
\end{figure}

Figures \ref{fig:decay_curves_dist5} and \ref{fig:decay_curves_dist7} show the decay curves for the $d=5$ and $d=7$ color codes, similar to the $d=3$ decay curves shown in figure \ref{fig:decay_curves_dist3} in the main text.

\begin{thebibliography}{99}
\bibitem{syndrome_increment}The syndrome increment is usually $\delta \vec{s}(t) \equiv \vec{s}(t) - \vec{s}(t-1) \mod 2$. When ancilla qubits are not reset between QEC cycles, we use a somewhat different definition, see App.~\ref{app:measurement_processing} for details.
\end{thebibliography}
\end{document}